\documentclass[3p,times,procedia]{elsarticle}
\usepackage{nupha_ecrc}
\usepackage[section]{placeins} 
\usepackage{amsthm}
\usepackage[font={small}]{caption}
\usepackage{mathrsfs}
\usepackage{graphics,fancyhdr,ifthen,amsmath,amssymb,color,graphicx,subfigure}
\usepackage{tikz}
\usepackage{hyperref} 

\volume{00}

\firstpage{1}

\journalname{Nuclear Physics A}

\runauth{}

\jid{nupha}

\jnltitlelogo{Nuclear Physics A}

\usepackage{amssymb}

\usepackage[figuresright]{rotating}

\begin{document}

\begin{frontmatter}



\dochead{}

\title{Measurements of correlations of anisotropic flow harmonics in Pb--Pb Collisions with ALICE}


\author{You Zhou (on behalf of the ALICE Collaboration) \footnote{Email: you.zhou@cern.ch}}

\address{Niels Bohr Institute, University of Copenhagen, Blegdamsvej 17, 2100 Copenhagen, Denmark}

\begin{abstract}

We report the first measurements of the correlation strength between various anisotropic flow harmonics, using ALICE data. This correlation strength is characterized with multi-particle cumulants of mixed harmonics, which by construction depend only on the fluctuations of magnitudes of the anisotropic flow vectors.
A detailed comparison to theoretical model calculations, including HIJING, Monte Carlo Glauber and hydrodynamics is also presented. These studies further constrain initial conditions, the properties and the evolution of the system to be used in theoretical simulations of heavy-ion collisions.
\end{abstract}


\end{frontmatter}


\section{Introduction}
\label{intro}

The primary goal of ultra-relativistic heavy-ion collisions is to investigate the properties of the quark-gluon plasma (QGP), a state of matter whose existence under extreme conditions is predicted by quantum chromodynamics. 
Measuring the anisotropic flow is an important step towards this goal.
It is traditionally quantified with harmonics $v_{n}$ and corresponding symmetry planes $\Psi_{n}$ in the Fourier series decomposition of the particle azimuthal distribution, in a plane perpendicular to the beam direction~\cite{Voloshin:1994mz}.
The anisotropic flow $v_{n}$ (for 1 $\leq n \leq$ 6) have been measured at the CERN Large Hadron Collider (LHC)~\cite{ALICE:2011ab}. These measurements provide compelling evidence that strongly interacting matter appears to behave like an almost perfect fluid~\cite{Heinz:2013th}. 
In addition to traditional anisotropic flow studies, it was realized that the fluctuations of the flow-vectors, whose magnitudes and directions are $v_{n}$ and $\Psi_{n}$, respectively, can bring new information about initial conditions and the properties of the QGP~\cite{Zhou:2014bba, Heinz:2013bua, Zhou:2015eya}. 
Recently, it is proposed to measure the Symmetric 2-harmonic 4-particle Cumulants, $SC(m,n)$, to study the relationship between different orders of anisotropic flow harmonics $v_m$ and $v_n$~\cite{Bilandzic:2013kga}. This new observable is defined as:
\begin{equation}
\begin{aligned}
SC(m,n) &~ = \left< \left< \cos (m\phi_{1} + n\phi_2 - m\phi_3 - n\phi_4) \right> \right>  -  \left< \left< \cos (m\phi_{1} - m\phi_2) \right> \right> \,\left< \left< \cos (n\phi_{1} - n\phi_2) \right> \right> \\
&~ = \langle v_m^2  \,  v_n^2 \rangle -  \langle v_m^2  \rangle \, \langle v_n^2 \rangle
\end{aligned}
\end{equation}
with $n \neq m$. 
By construction, this observable is not sensitive to either non-flow effects due to usage of 4-particle cumulant, or inter-correlations of various symmetry planes. It is non-zero if there is correlation (or anti-correlation) between $v_m$ and $v_n$.

Here we present the measurements of $SC(4,2)$ and $SC(3, 2)$ in Pb--Pb collisions at $\sqrt{s_{\rm NN}} = $ 2.76 TeV in ALICE. The comparisons of experimental measurements and calculations from the HIJING model, Monte Carlo Glauber (MC-Glauber) and hydrodynamic calculations will also be discussed.

\section{Analysis Details}

The data samples collected by ALICE in the first Pb--Pb run at the Large Hadron Collider were used in this analysis. For more details about the ALICE detector, refer to~\cite{Aamodt:2008zz}. About 16 million Pb--Pb events were recorded with a minimum-bias trigger, based on signals from two VZERO detectors (-3.7$\textless \eta \textless$-1.7 for VZERO-C and 2.8$\textless \eta \textless$5.1 for VZERO-A) and from the Silicon Pixel Detector. Charged particles are reconstructed using the Inner Tracking System and the Time Projection Chamber with full azimuthal coverage in the pseudo-rapidity range $|\eta|\textless$0.8. 
The 2- and multi-particle correlations are measured using the generic framework introduced in~\cite{Bilandzic:2013kga}.

\section{Results and Discussions}


\begin{figure}[tbh]
\begin{center}
\includegraphics*[width=7cm]{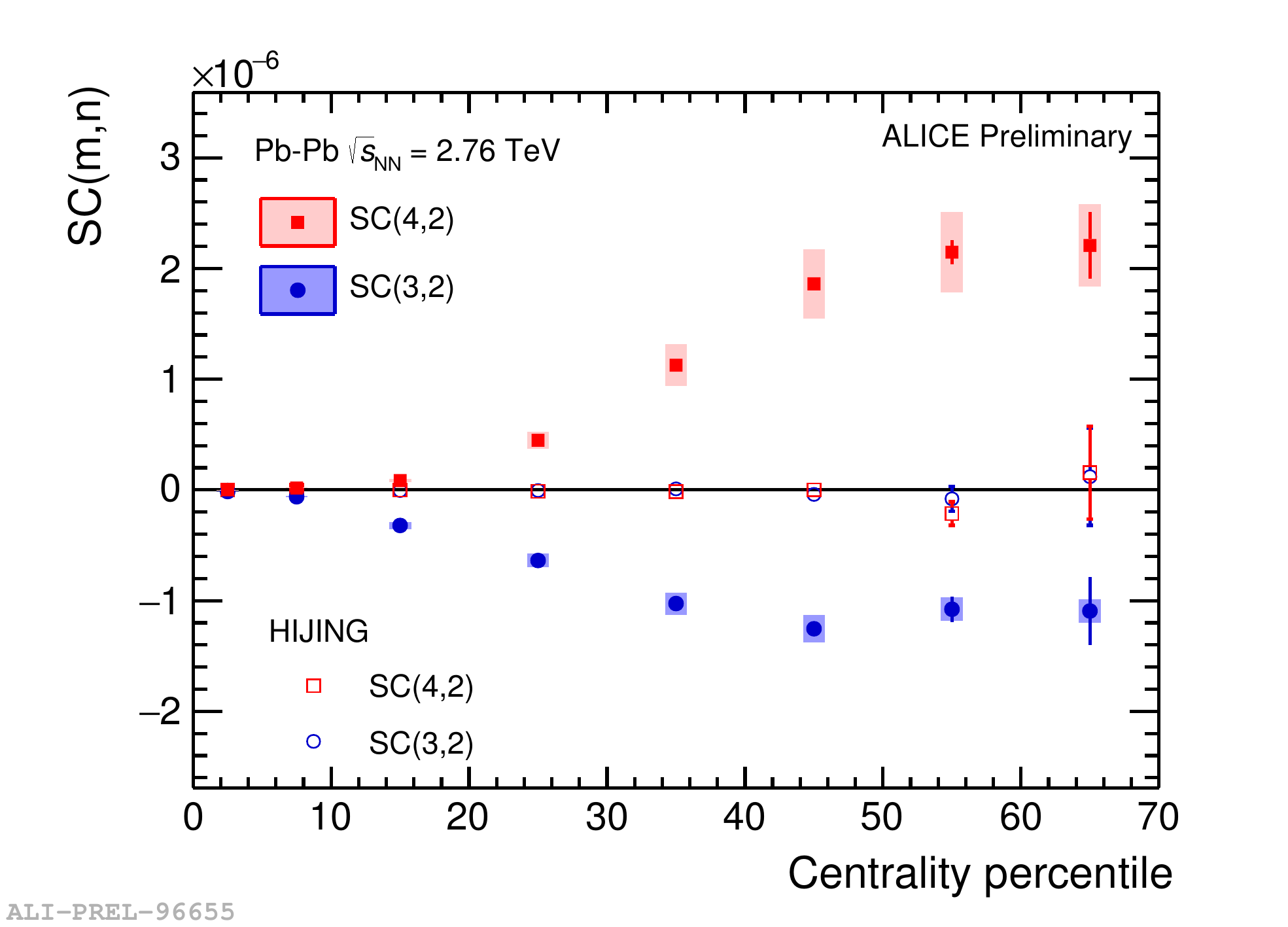}
\includegraphics*[width=7cm]{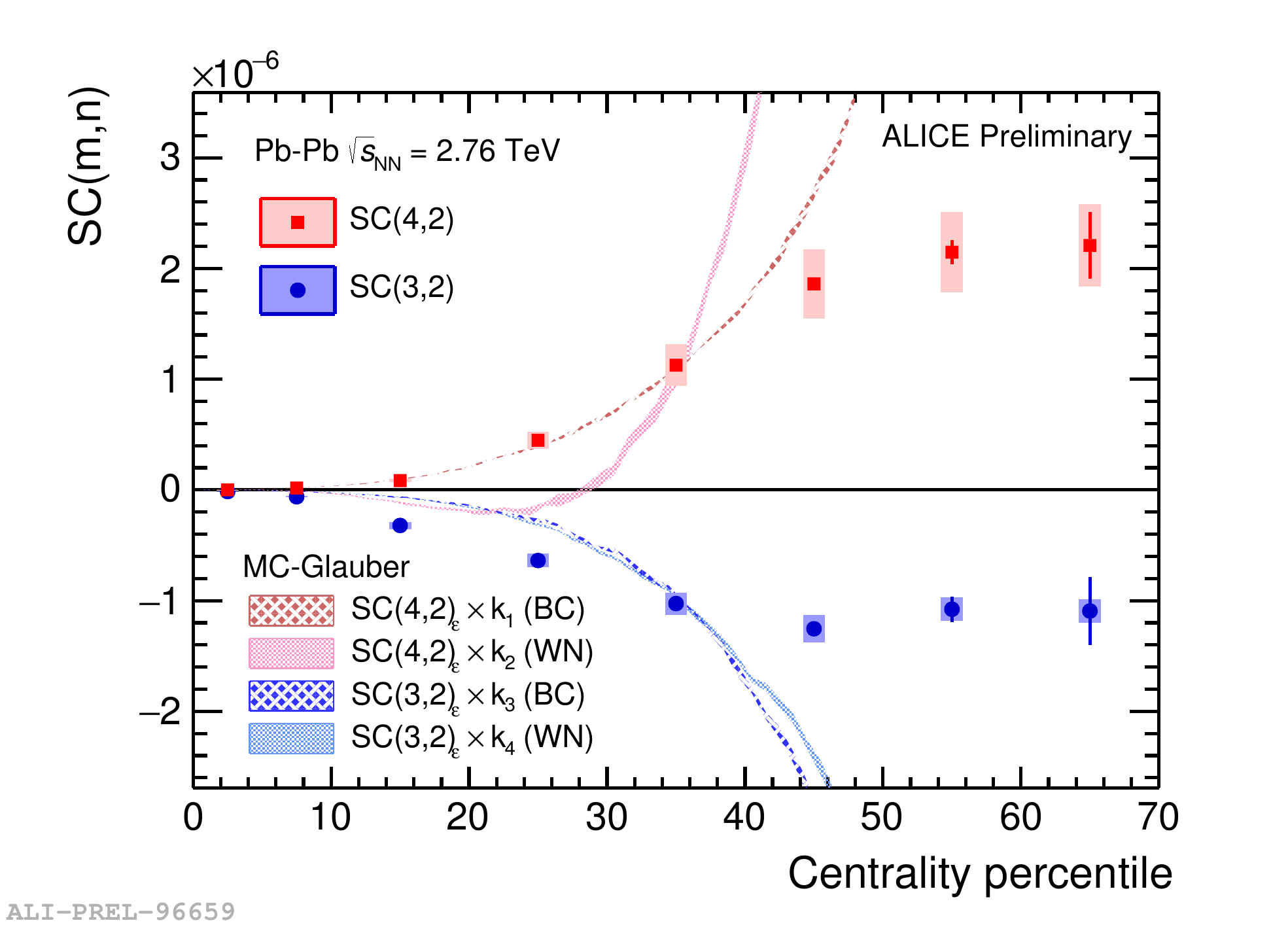}
\caption{
(Left) Centrality dependence of $SC(4,2)$ (red solid squares) and $SC(3, 2)$ (blue solid circles) in Pb--Pb collisions at 2.76 TeV. Results for HIJING model are also presented here. (Right) The comparison of SC measurements to the MC-Glauber calculations. 
}
\label{fig:f1}
\end{center}
\end{figure}

Figure~\ref{fig:f1} (left) presents the centrality dependence of $SC(4,2)$ (red solid squares) and $SC(3, 2)$ (blue solid circles) for Pb--Pb collisions at 2.76 TeV. Positive values of $SC(4,2)$ and negative $SC(3,2)$ are observed for all centralities. This shows a correlation between $v_2$ and $v_4$, and an anti-correlation between $v_2$ and $v_3$. These results suggest that finding $v_2$ larger than $\left< v_2 \right>$ in an event enhances the probability of finding $v_4$ larger than $\left< v_4 \right>$, and enhances the probability of finding $v_3$ smaller than $\left< v_3 \right>$ in that event. The calculations of $SC(4,2)$ and $SC(3, 2)$ using HIJING~\cite{Wang:1991hta}, which does not include anisotropic collectivity, are compared to the experimental measurements. We found that both 4-particle correlation $\left< \left< \cos (m\phi_{1} + n\phi_2 - m\phi_3 - n\phi_4) \right> \right> = \langle v_m^2  \, v_n^2 \rangle$ and the product of the 2-particle correlation $\left< \left< \cos (m\phi_{1} - m\phi_2) \right> \right> \,\left< \left< \cos (n\phi_{1} - n\phi_2) \right> \right> =  \langle v_m^2  \rangle \, \langle v_n^2 \rangle$ are non-zero using HIJING simulations. However, the HIJING calculations show that both $SC(4,2)$ and $SC(3, 2)$ are consistent with zero. This suggests that the observable SC is nearly insensitive to non-flow effects. Non-zero values of $SC(4,2)$ and $SC(3, 2)$ can not be explained by non-flow effects, and confirms the existence of correlations (and anti-correlations) between anisotropic flow harmonics.

It can be seen that the correlation strength of the measured SC increases non-linearly up to centrality 60\%. We therefore investigate to what extent this non-trivial trend can be captured solely by correlations of corresponding anisotropies in the initial state.
The centrality dependence of $SC(4, 2)$ and S$C(3, 2)$ from two scenarios of MC-Glauber simulations, with Wounded Nucleon (WN) and Binary Collisions (BC) weight, are also compared to SC measurements in Fig~\ref{fig:f1} (right). The correlations of $m^{th}$ and $n^{th}$ order anisotropies, $\varepsilon_{m}$ and $\varepsilon_{n}$, were estimated with calculations of $\langle \varepsilon_{m}^{2} \, \varepsilon_{n}^{2} \rangle - \langle \varepsilon_{m}^{2} \rangle \, \langle \varepsilon_{n}^{2} \rangle$ and scaled to match the 30- 40\% data using scaling number $k_{n}$. Here the $\varepsilon_{m}$ (or $\varepsilon_{n}$) is the anisotropy in coordinate space.
Increasing trend from central to peripheral collisions with different signature has been observed for $SC(4, 2)$ and $SC(3, 2)$, however the centrality dependence of corresponding observables measured in momentum space cannot be captured well. This indicates that the relationship between different order anisotropies is not the only contribution to observed (anti-)correlations of anisotropic flow harmonics $v_{m}$ and $v_{n}$.

\begin{figure}[tbh]
\begin{center}
\includegraphics*[width=10cm]{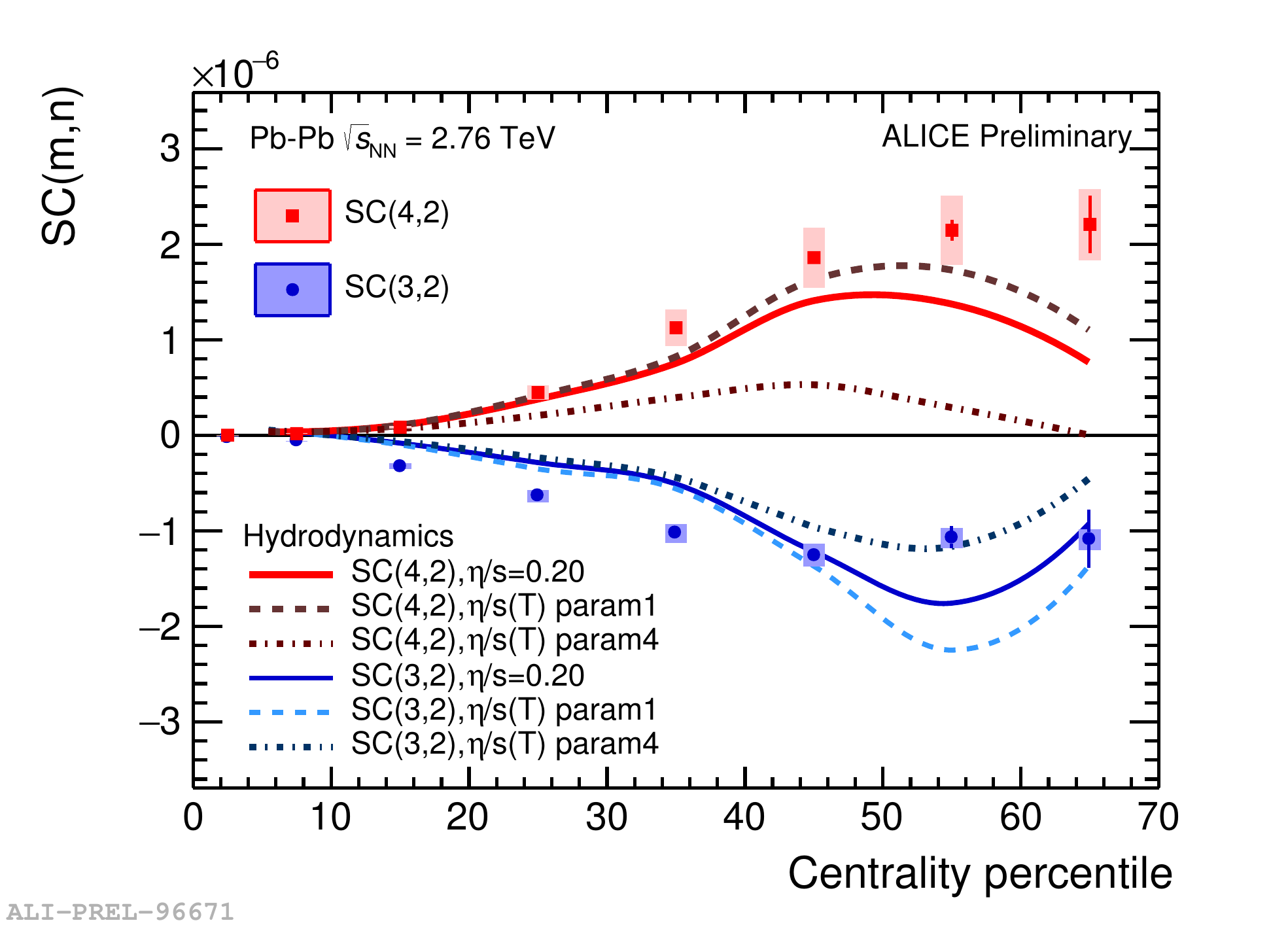}
\caption{Centrality dependence of $SC(4,2)$ and $SC(3,2)$ measured in Pb--Pb collisions at 2.76 TeV and the comparisons to hydrodynamic calculations using various $\eta/s$(T) parameterizations from~\cite{Niemi:2015qia}. }
\label{fig:f2}
\end{center}
\end{figure}

It was shown in AMPT model calculations that the observed (anti-)correlations are also sensitive to the transport properties of the created system, e.g. the partonic and hadronic interactions~\cite{Zhou:2015eya, Bilandzic:2013kga}. To further study the properties of QGP, we show in Fig.~\ref{fig:f2}  the comparison of SC measurements and hydrodynamic model calculations, which incorporate both initial conditions and the dynamic evolution. It is observed that SC is very sensitive to the values of $\eta/s$ in this model. Thus, the investigation of SC provides a new approach to constrain the input of $\eta/s$ in hydrodynamic calculations, in addition to standard anisotropic flow studies.
We also notice that although this hydrodynamic model reproduced the centrality dependence of $v_n$ for $n \leq 2$ fairly well using various $\eta/s$(T) parametrization, its predictions of SC cannot quantitatively describe the data. There is no single centrality for which a given $\eta/s$(T) parameterization describes simultaneously $SC(4,2)$ and $SC(3,2)$. Therefore, it is believed that together with the individual flow harmonics, the new observable SC provides stronger constrains on the $\eta/s$ and initial conditions than standard $v_n$ measurement alone.

\begin{figure}[tbh]
\begin{center}
\includegraphics*[width=10cm]{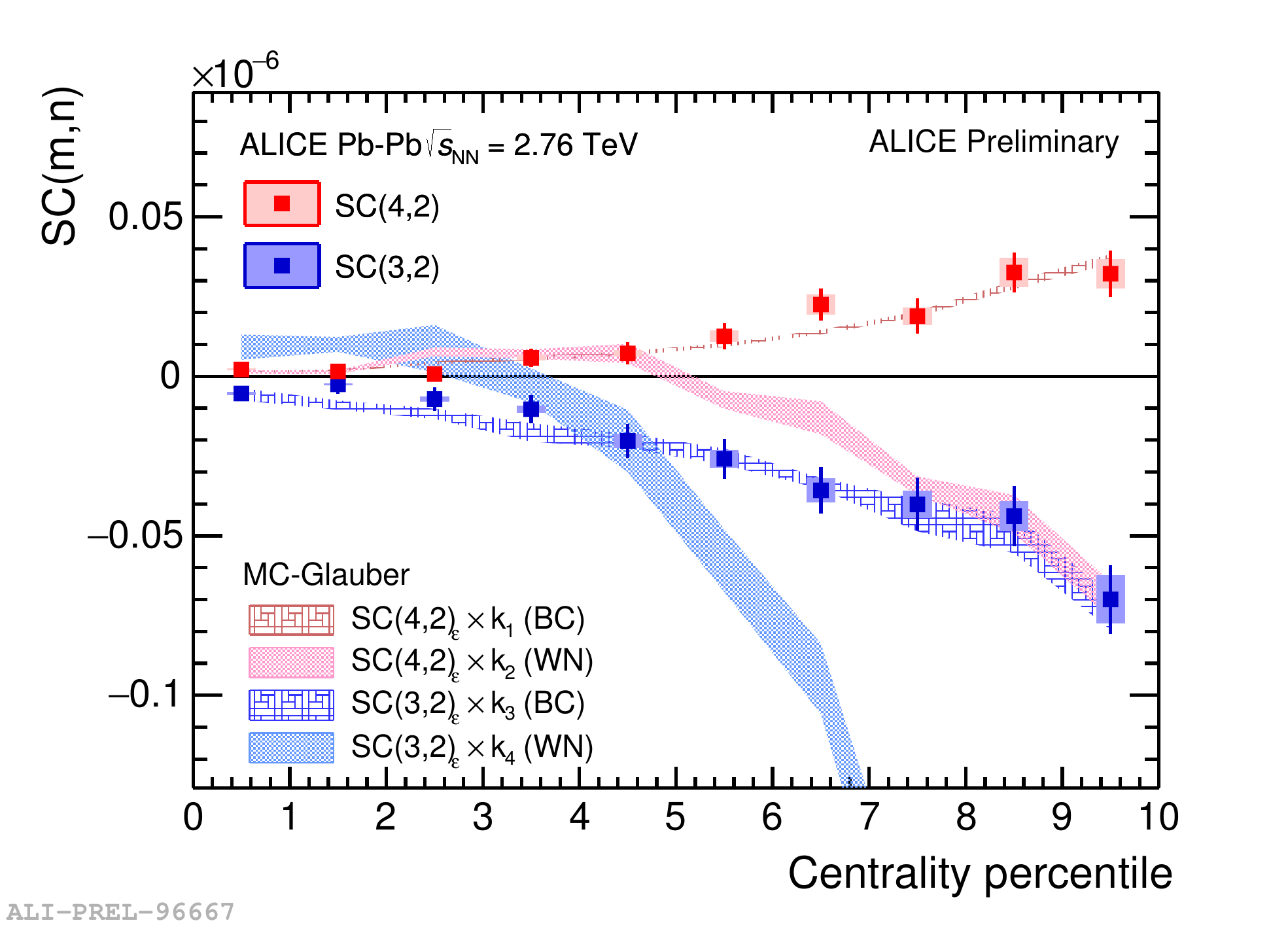}
\caption{ Centrality dependence of $SC(4,2)$ and $SC(3,2)$ for 0-10\% central Pb--Pb collisions, MC-Glauber calculations have been scaled to match the 4-5\% data using scaling number $k_{n}$.
}
\label{fig:f3}
\end{center}
\end{figure}

Figure~\ref{fig:f3} shows the SC measurements in 0-10\% central collisions where the anisotropies originate mainly from initial geometry fluctuations. The non-zero values of SC are observed, and $v_{2}$ and $v_{4}$ remain correlated, $v_{2}$ and $v_{3}$ anti-correlated. Assuming $v_{n} \propto \varepsilon_{n}$ in the central collisions, we expect the normalized SC from initial conditions should follow the trend of SC measurements. The results show that the MC-Glauber calculations using BC parameterization is favored by the data in this region.

\section{Summary}

In summary, we present for the new measurements Symmetric 2- harmonic 4-particle Cumulants, $SC(m,n)$, which quantify the relationship between $m^{th}$ and $n^{th}$ order flow harmonics $v_m$ and $v_n$. 
Positive $SC(4,2)$ and negative values of $SC(3,2)$ are observed for all presented centralities. Since this observable is insensitive to non-flow effects, as suggested by the study of simulated HIJING events, the above results suggest an intrinsic correlation between $v_{2}$ and $v_{4}$, and an anti-correlations of $v_{2}$ and $v_{3}$. The comparison to MC-Glauber calculations and hydrodynamic calculations further suggests that the combined investigations of $v_{n}$ and $SC(m,n)$ provide a powerful handle to determine the initial conditions and temperature dependence of $\eta/s$ in ultra-relativistic heavy ion collisions.





\bibliographystyle{elsarticle-num}
\bibliography{<your-bib-database>}



\end{document}